\documentclass[lettersize,journal]{IEEEtran}
\usepackage{amsmath,amsfonts}
\usepackage{algorithmic}
\usepackage{algorithm}
\usepackage{array}
\usepackage[caption=false,font=normalsize,labelfont=sf,textfont=sf]{subfig}
\usepackage{textcomp}
\usepackage{stfloats}
\usepackage{url}
\usepackage{hyperref}
\usepackage{verbatim}
\usepackage{graphicx}
\usepackage{multirow}
\usepackage{cite}
\usepackage{color}
\begin{document}

\title{An Effective Learning Management System for Revealing Student Performance Attributes}

\author{Xinyu Zhang, \IEEEmembership{Member, IEEE,} Vincent CS Lee, \IEEEmembership{Senior Member, IEEE,} Duo Xu, Jun Chen,  \IEEEmembership{Member, IEEE,} Mohammad S. Obaidat, \IEEEmembership{Life Fellow, IEEE}

\thanks{Xinyu Zhang is with the School of Electronics and Information, Northwestern Polytechnical University, Xi'an $710072$, China. (email: xinyu.zhang@nwpu.edu.cn)}
\thanks{Vincent CS Lee is an Associate Professor at the Department of Data Science and Artificial Intelligence, Faculty of IT, Monash University, Melbourne $3800$, Australia. (email: vincent.cs.lee@monash.edu)}
\thanks{Duo Xu is with Guangyuan Foreign Language School, Guangyuan $628018$, China. (email: doris.xu1117@gmail.com)}
\thanks{Jun Chen is an Associate Professor with the School of Electronics and Information, Northwestern Polytechnical University, Xi'an $710072$, China; Chongqing Institute for Brain and Intelligence, Guangyang Bay Laboratory, Chongqing $400064$, China. (email: junchen@nwpu.edu.cn)}
\thanks{Mohammad S. Obaidat is a Distinguished Professor at the King Abdullah II School of Information Technology, The University of Jordan, Amman $11942$, Jordan and School of Computer and Communication Engineering, University of Science and Technology Beijing, Beijing $100083$, China, Department of Computational Intelligence, School of Computing, SRM University, SRM Nagar, Kattankulathur $603203$, TN, India and School of Engineering, The Amity University, Noida, UP $201301$, India. (email: msobaidat@gmail.com or m.s.obaidat@ieee.org)}
}

\markboth{}%
{Shell \MakeLowercase{\textit{et al.}}: A Sample Article Using IEEEtran.cls for IEEE Journals}


\maketitle

\begin{abstract}
Contribution: A learning management system \emph{(LMS)} incorporated with an advanced educational data mining module is proposed, as a means to mine efficiently from student performance records to provide valuable insights for educators in helping plan effective learning pedagogies, improve curriculum design, and guarantee quality of teaching.

Background: An LMS streamlines the management of the teaching process in a centralized place, recording, tracking, and reporting the delivery of educational courses and student performance. 
Educational knowledge discovery from such an e-learning system plays a crucial role in rule regulation, policy establishment, and system development. 
However, existing LMSs do not have embedded mining modules to directly extract knowledge.
As educational modes become more complex, educational data mining efficiency from those heterogeneous student learning behaviours is gradually degraded.

Intended outcomes: The design and application of the LMS enable educators to learn from past student performance experiences, empowering them to guide and intervene with students in time, and eventually improve their academic success.

Application design: This study proposes an effective LMS which processes the stored data through an advanced educational data mining module. 
The mining module utilizes probability-based frequent pattern mining for generating candidate itemsets and updating LMS databases in each iteration for complexity reduction. 
In addition, the conditional probability increment ratio (\emph{CPIR}) is deployed as an exceptionality measure to mine effective common and exception rules simultaneously.
The iteratively generated rules can reveal academic performance patterns so that have merits in learning personal goal achievement. 

Findings: Through two illustrative case studies, experimental results demonstrate increased mining efficiency of the proposed mining module without information loss compared to classic educational mining algorithms. 
The mined knowledge reveals a set of attributes that significantly impact student academic performance, and further classification evaluation validates the identified attributes. 
\end{abstract}


\section{Introduction}
\IEEEPARstart{L}{earning} management systems (\emph{LMSs}) are designed to foster student engagement and initiative, creating an atmosphere where learners can effectively work on their own development and learning journey \cite{Ogu2021}.
LMS is an integrated learning platform that stores, tracks, and reports information, such as course materials, communications, instructions, and performance evaluations.
Educational data mining from such a system is crucial in educational sectors to enable teaching staff to adapt dynamic teaching strategies and positively influence students' progressive development and education quality policy-making, thereby fostering the establishment of well-behaved study habits.
However, mining from an LMS is challenging, as the system has heterogeneous records representing student learning behaviors, including varied cognitive styles, learners' emotional behaviours, and learners' personal goals.

In this context, association rule mining (\emph{ARM}), is a subset of data mining techniques, which is employed to uncover hidden correlations among various attributes of student progress and performance \cite{Ko2020}. 
These correlations provide valuable insights that can be translated into actionable knowledge in the educational domain.
Specifically, the knowledge extracted from ARM techniques proves instrumental in the exploration of learning analytics \cite{Ald2019}, facilitating policy establishment \cite{Zhang2022}, and developing educational systems \cite{Tel2020}. 
These applications highlight the significance and feasibility of utilizing ARM techniques to predict students' academic success.

The concept of ARM, introduced by Agrawal et al. \cite{Agr1993}, brought widespread applications of the Apriori algorithm to extract valuable information. 
The Apriori algorithm identifies frequent candidate itemsets based on two measurements: the support represents the frequency of the generated rules, and the confidence indicates the reliability of the generated rules.
However, the Apriori algorithm has certain limitations. 
One of the main challenges is its need to scan the entire original database in each rule generation iteration to produce candidate frequent itemsets. 
As a result, the computational requirements of this algorithm increase exponentially when dealing with large and complex datasets.
Furthermore, the patterns extracted by the Apriori algorithm can be grouped into three types: common rules (i.e., high frequency and high reliability), reference rules (i.e., low frequency and low reliability), and exception rules (i.e., low frequency but high reliability) \cite{Dav2008}. 
Previous studies majorly focus on deriving common rules as they interpret the regularity of objects with consequences, whereas reference and exception rules were generally discarded.
However, exception rules hold the potential for being more engaging, significant, and valuable than common rules \cite{Mahm2022}. 
This is because they can provide information that reveals unusual and contradictory but knowledgeable insights. 
Accordingly, considering the above limitations and exploring an advanced ARM technique is critical to offer valuable perspectives in analyzing student performance patterns from a complex LMS.

Nevertheless, existing studies generally retrieve LMS records for educational mining as an external step, which cannot extract knowledge iteratively, and this aggregates the gap between learning behaviors and policy establishment.
To address these challenges, this study proposes an effective LMS embedding with an advanced ARM module consisting of a proposed Faster Apriori algorithm. 
The proposed ARM module deploys probability-based frequent pattern mining approach to generate candidate itemsets and introduces an updated database during each rule generation iteration, which excludes items not shortlisted for dimensional reduction.
Additionally, the proposed Faster Apriori algorithm utilizes the conditional-probability increment ratio (\emph{CPIR}) \cite{WuX2004} as an additional rule measurement to filter out reference rules, thereby retaining only common and exception rules for valuable knowledge extraction, and this guarantees meaningful information to be preserved.
The proposed system was applied to process educational records, revealing valuable student performance patterns that can guide their academic success.
In summary, the key contributions can be outlined as follows:

\begin{itemize}
\item  Taking account of computational complexity, hybrid mode of teaching and availability of socio-economic resources, this study proposes an effective LMS that aims to unravel academic performance pattern for learning personal goal achievement. 

\item An advanced educational mining module is introduced and embedded in the proposed LMS. 
The proposed Faster Apriori algorithm is designed to extract rules more efficiently.
The algorithm incorporates probability-based frequent pattern mining with overriding original databases to repel under-qualified itemsets in each interaction for dimensional reduction.
This algorithm guarantees increased efficiency without information loss.

\item The proposed algorithm generates information by employing the CPIR measurement, so that the reference rules will be filtered out, guiding valuable knowledge discovery with common and exception rules in place.

\item This study reveals attributes contributing to student performance, which can be applied to assist their academic success. 
This knowledge enables targeted interventions and personalized educational planning, optimizing support for students and fostering their effective study habits.

\item This work ensures high reproducibility by utilizing open-access datasets for evaluation. 
The proposed algorithm is fully accessible to the research community through the GitHub repository: \url{https://github.com/Amyyy-z/Association-rule-mining}.
\end{itemize}

\section{Literature Review}
There is a rapid growth in research using ARM for LMS development \cite{Can2019, Hus2019, Czi2019, Roja2019, Moh2023}. 
This section reviews some studies on ARM applications in education and explains their limitations.

\subsection{ARM in Educational Research}
Cantabella et al. \cite{Can2019} once applied ARM through a big data framework to accelerate the statistical analysis of students' learning behaviors in a LMS.
Hussain et al. \cite{Hus2019} deployed the Apriori algorithm to analyze $666$ instances with $11$ attributes, and results suggested that mothers' education and coaching classes are correlated with students' performance.
Czibula et al. \cite{Czi2019} applied the relational association rules (\emph{RARs}) and classification to predict students' final results.
The RARs were introduced in \cite{Ca2006}, which were used to identify the non-ordinal relation between attributes.
Accuracies were achieved through three distinct datasets: $0.79$, $0.70$, and $0.81$.
Similarly, Rojanavasu \cite{Roja2019} combined the Apriori algorithm with a decision tree classifier to predict potential jobs after graduation.
The results indicated that region, admission project, faculty, and province are considerably related to students' future careers.  
Mohamad and Tasir \cite{Moh2023} utilized ARM on qualitative data of reflections to model the impact of questioning-based feedback on reflective thinking skills.
The results indicated that despite higher-level student's feedback, their reflective discussions tend to remain limited to descriptive thinking and decision-making.
Other studies proved the benefits of applying ARM techniques in educational data mining \cite{Ko2020, Rom2020, Mar2020, Ya2022, Imh2022, And2023, Bat2023, Kumar2023}.

\subsection{ARM Research}
The continuous development of more advanced ARM algorithms have enabled researchers to effectively extract valuable insights from complex datasets.  
For example, Tao et al. \cite{TaoF2003} introduced weighted ARM, which assigns a weighted support value to improve rule selection. 
Buczak et al. \cite{Buc2010} proposed fuzzy ARM to address the uncertainty of numerical attributes by incorporating fuzzy logic. 
Other studies have focused on enhancing the validity of generated rules from a confidence perspective, including fuzzy weighted ARM \cite{Muy2009} and profile-based fuzzy ARM \cite{Yav2021}.
While there is a significant body of existing literature addressing the effectiveness of ARM applications, there seems to be a lack of studies focusing specifically on the efficiency of these techniques.

Several studies proposed advanced ARM algorithms to incorporate different data structures to address the efficiency issue arising from the classic candidate itemsets generation through table-based procedures. 
Specifically, the tree-based structure is the most rational and frequently adopted type \cite{Han2000, Fern2001}.
Moreover, Le et al. \cite{LeHo2018} introduced an animal migration optimization algorithm, which can address the issue of efficient rules generation. 
Lin et al. \cite{LinW2002} proposed an ARM algorithm that selects rules based on a specified target range or number, further contributing to efficiency improvements.
Indeed, tree-based structures may become less computationally efficient when dealing with high-dimensional and complex datasets. 
As the dimensionality or volume increases, the process of generating the tree structure for mining explicit knowledge (rule-based representation) can become more time-consuming and resource-intensive. 

Extracting intriguing and unexpected associations, in addition to enhancing the performance of ARM techniques, is indeed a challenging research direction.
However, existing studies have often overlooked the extraction of exception rules, which can provide valuable insights.
In works such as \cite{Dav2008} and \cite{Daly2004}, an exceptionality measure was proposed to select meaningful exception rules. 
Extraction of exception rules is crucial as they can contribute to systems design, policy establishment, decision-making processes, and identifying and preventing risk factors. 
Incorporating the extraction of exception rules in ARM techniques can uncover valuable and actionable knowledge that goes beyond regular patterns, which is especially vital in educational data mining to enhance the quality of education and reduce potential failure rates.

Collectively, in the realm of education data mining, one of the primary research objectives is to enhance the efficiency of ARM techniques, particularly when dealing with high-complexity LMS records with large volumes of iterative feedback.
Additionally, the analysis of exception rules, which have been largely overlooked by existing studies, is of utmost importance. 
By addressing both efficiency challenges and the exploration of exception rules, education data mining can unlock a wealth of actionable knowledge for improving educational outcomes.

\section{Methodology}
With the objective of efficiently extracting valuable knowledge from massive educational data, the proposed LMS architecture and the embedded ARM module are explained here.

\subsection{Proposed Effective LMS}
\begin{figure}[!t]
\centering
\includegraphics[width=0.8\columnwidth]{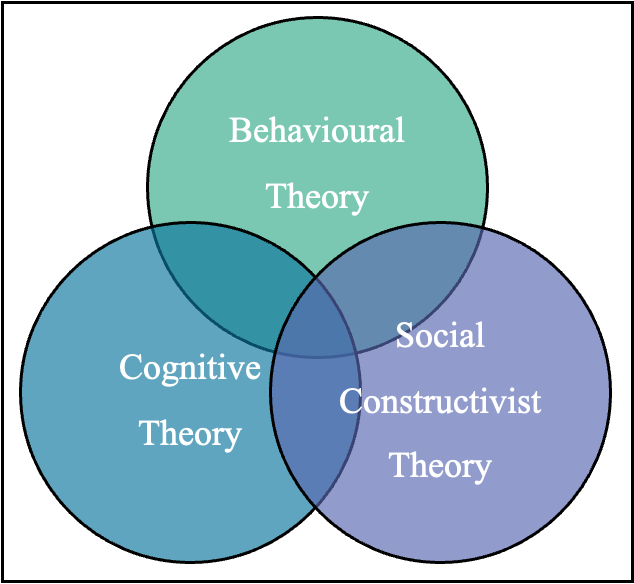}
\caption{Composite Theory (Adapted from Trakunphutthirak and Lee \cite{Trak2022}).}
\label{fig:fig1}
\end{figure}

Bryn and Gardner \cite{Bryn2006} once stated that there were three critical theoretical concepts related to the effective development of an LMS, including behavioral, cognitive, and social constructivist theories.
The behavioral theory highlights the interaction between a stimulus and a response, and it indicates the significance of a supporting LMS which determines students' behaviours.
The cognitive theory emphasises the interaction between external and internal stimuli, which uses a learning system to support and enhance cognition.
The social constructivist theory focuses on the interaction between students and the others in a collaborative learning atmosphere.
Those three theories overlap among one and the other, suggesting that no single one of them can solely be a holistic approach to explain or develop an effective LMS.
Accordingly, this study takes our pioneer study conducted by Trakunphutthirak and Lee \cite{Trak2022} in $2022$, which proposed the composite theory (see Fig. \ref{fig:fig1}) for LMS construction.

The composite theory integrates students' behaviours, cognition, and social interaction, which are stored and accessed through a holistic LMS archival. 
The overlapping and complementary components of the theory can be easily captured through the LMS \cite{Trak2022}.
Lemantara et al. \cite{Lem2023} mentioned that there are additionally three laws of learning, including the law of readiness, the law of exercise, and the law of effect \cite{Ni2020}, and those learning theories can be achieved through the three dimensions of the composite theory.
Inspired by their work, this study utilises the composite theory proposed in our pilot study to shape an effective LMS with an embedded data mining module to store and mine essential features that affect student performance success.
Under such a scenario, this study aims to reach precision education for practitioners and also policymakers in the effective implementation and development of valuable knowledge mining.

\begin{figure}[!t]
\centering
\includegraphics[width=1.0\columnwidth]{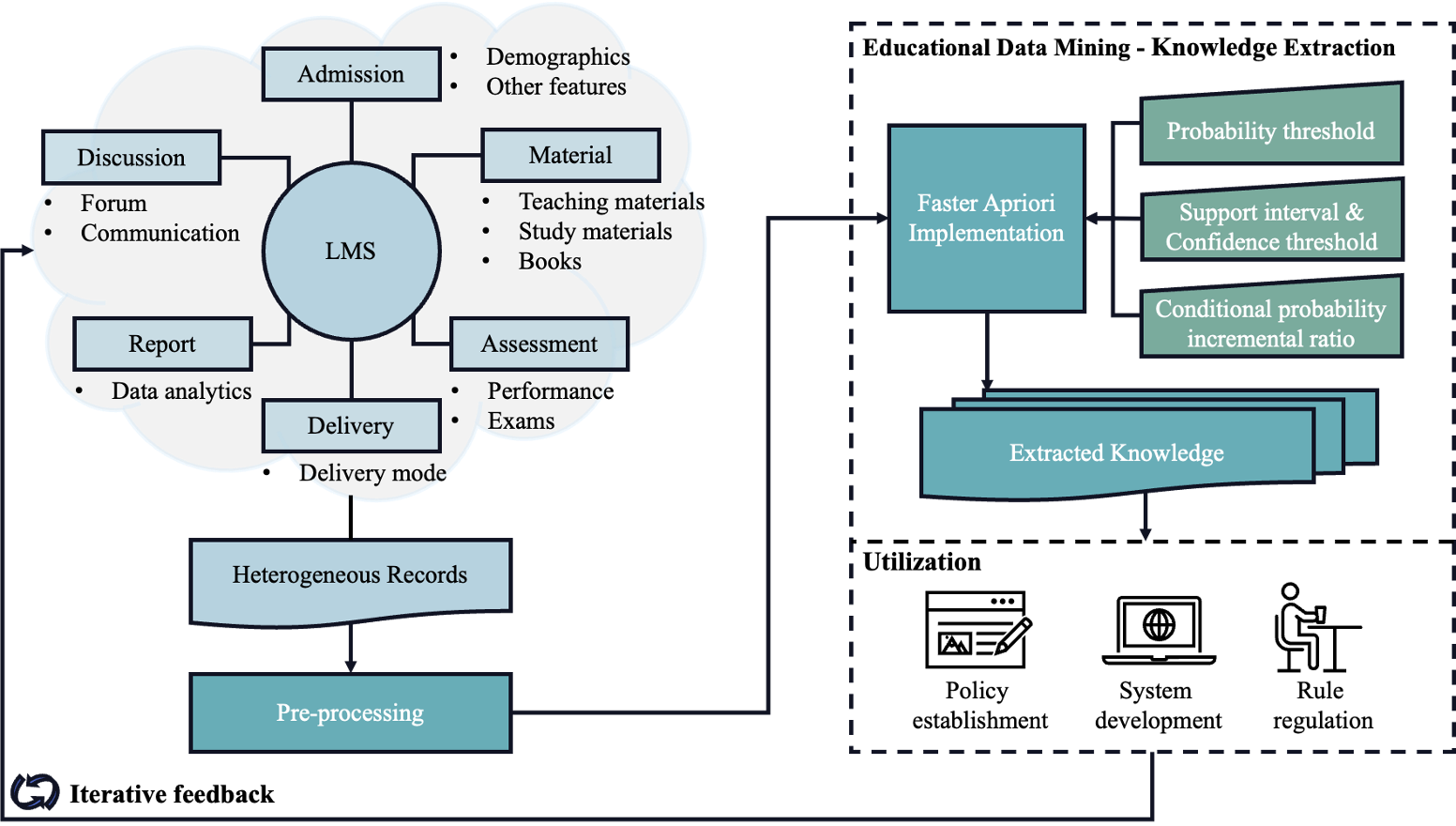}
\caption{Proposed Effective LMS Framework.}
\label{fig:fig2}
\end{figure}

Fig. \ref{fig:fig2} exhibits the proposed effective LMS framework.
In daily educational activities, LMS platforms store and generate a substantial number of records, encompassing students' learning materials, discussions, performance evaluations, and more. 
These heterogeneous records often exist in semi-structured or mostly unstructured formats, holding the potential for extracting intricate knowledge that can bring innumerable benefits.
Existing LMS usually stand alone without the processing and mining phases for interacting with those stored data \cite{Tran2020, Ramos2021}.
However, these studies neglect to extend LMS to a consolidated system for efficiently mining from records for knowledge discovery.
Therefore, our framework introduces a novel ARM module, powered by the proposed Faster Apriori algorithm, which efficiently mines valuable knowledge (i.e., common and exception rules) simultaneously, while adhering to pre-defined probability, support, confidence, and CPIR thresholds. 
The extracted knowledge can be incorporated iteratively into the LMS design, influencing policy-making, systems development, and rules regulation to enhance the overall learning experience and outcomes.

\subsection{Proposed ARM Module - Faster Apriori}
The proposed ARM module includes a major algorithm component, named the Faster Apriori algorithm, which utilizes a probability-based frequent pattern mining approach to generate candidate frequent itemsets.
It initially identifies the occurrence of each items by estimating their frequent probabilities, which has been proven effective in \cite{Rong2012}.
Then, the algorithm improves rule extraction efficiency by overriding the original database during each frequent itemset generation iteration by removing itemsets that are below the pre-defined threshold.
This approach can reduce computational complexity while enhancing efficiency without sacrificing information.
Additionally, this study incorporated the CPIR \cite{WuX2004} exceptionality measurement to select high-quality common and exception rules simultaneously, enabling the discovery of more valuable and meaningful knowledge.

The extracted knowledge from the Faster Apriori algorithm is to verify $X \Rightarrow Y$ indicating if the item $X$ exists as ``Antecedence'', then item $Y$ should co-exist as ``Consequence''.
Therefore, the original LMS student learning behaviors database can be denoted as $D = \{X_{1}^{m}, X_{2}^{m}, \dots, X_{n}^{m}\}$ where $X_i$ is the $i$th instance in $D$ with multiple attributes $m$, and $n$ is the total number of instances $X$ in $D$.
The initial $D$ will be utilized to identify unique items of $X_{n}^{m}$ by identifying the probability of individual unique $X_{n}^{m}$ attribute, in this case, the critical features affecting students' performance.
The probability of combinations for the unique $X_{n}^{m}$ will also be calculated and compared to the pre-defined threshold (i.e., support, confidence, and CPIR).
Following the exclusion of under-qualified $X_{n}^{m}$ and the corresponding combination pairs, a new dataset $\bar{D}$ will be generated for frequent itemsets generation intuitively.
To further reduce the computational cost for efficiency optimization, the original dataset $D$ will be updated to $\bar{D}$ in each iteration when finding frequent itemsets to exclude redundant itemsets.

Algorithm \ref{algorithm} demonstrates the detailed procedure of the proposed Faster Apriori.
Before generating candidate itemsets, the probability-based frequent patterns will be generated by utilizing the probability measure $P_X$ to identify the probability scores for all unique items in $X$, in other words, to identify all relevant factors that affect students' performance.
The $P$ score will be measured to evaluate all the possible combinations with unique $X_{n}^{m}$, and this is to find all possible combinations of the qualified factors.
Generated unique $X_{n}^{m}$ and its corresponding combinations will be compared to the pre-defined probability threshold, and then the under-qualified ones will be removed, whereas the qualified ones will be remained to generate a new $\bar{D}$ for overriding the original $D$.
This process eliminates the itemsets that are not frequently seen in the database, thus reducing the running time in the later iterations for frequent candidate itemsets generation.
The support values will be calculated for all the newly generated candidate itemsets.
During each iteration, the database $\bar{D}$ will be updated to exclude those below the threshold.
In this regard, the original database will be overridden to reduce computational needs in each iteration, and the meaningless itemsets will not be considered for rules generation, thus enhancing the efficiency to an optimized level.
Then, the evaluation of the frequent itemsets after looping through all the iterations will be the confidence and CPIR values to further select the most valuable rules, in the meantime, would be the most significant factors.

\begin{algorithm}[!t]
\caption{Faster Apriori with Exceptionality.}\label{algorithm}
\begin{algorithmic}
\STATE $D = X_{n\in(0, N)}^{m(\geq 1)}$; \\
\STATE $\theta_{Prob}, \theta_{Sup}, \theta_{Con}, \theta_{CPIR}$; Threshold values \\

\STATE Generate initial $item_{ini}$ with all unique items in $X$ \\
\STATE Calculate $P_X$ for all $item_{ini}$ with Eq. \ref{eq:px} \\
\STATE Calculate $P$ for paired combinations $item_{pair}$ of items in $item_{ini}$ with Eq. \ref{eq:prob}\\
\STATE Exclude $item_{ini}$ and $item_{pair}$ below $\theta_{Prob}$ \\
\STATE Produce candidate itemsets $C^m$ with remaining $item_{ini}$ and $item_{pair}$ \\
\STATE Generate $\bar{D}$ based on $C^m$ \\

\STATE \textbf{Initialization:} \\
\STATE Set $n = 0$ \\
\STATE Set $itemset_{temp}$ to an empty list \\

\WHILE {$n < \bar{N}$}
    \STATE Set $i = 1$;  $i$ is the number of $i$ items in $\bar{X}_n$ \\
    \WHILE {$i \leq len(\bar{X}_{n}^{m})$}
        \IF {$i$ not in $C^m$}
            \STATE $i$ append to $C^m$ \\
        \ENDIF
        \STATE Calculate support values for all $C^m$ using Eq. \ref{eq:sup} 
        \IF {$Sup_{i} > \theta_{Sup}$}
            \STATE $i$ append to $itemset_{freq}$; 
            \STATE Update $\bar{D}$ with $itemset_{freq}$
        \ENDIF \\
        $i = i + 1$ \\
    \ENDWHILE \\
    $n = n+1$ \\
\ENDWHILE

\STATE Calculate confidence for all $itemset_{freq}$ use Eq. \ref{eq:con} \\
\STATE Calculate CPIR for all $itemset_{freq}$ use Eq. \ref{eq:com} \\
$Rules \gets itemset_{freq}, Sup_i, Con_i, CPIR_i$ \\
\IF {$Rules \geq \theta_{Sup}, \theta_{Con}, \theta_{CPIR}$}
    \STATE $Rules$ will be stored
\ENDIF
\STATE Store and plot all the final rules \\
\STATE {\bf Output:} $Final$ $\gets$ $Rules$ $(Sup, Con, CPIR)$; store all qualified rules with exceptions
\end{algorithmic}
\end{algorithm}

\subsection{Evaluation Metrics}
To extract valuable information, four metrics were employed: probability, support, confidence, and CPIR to generate final rules.
Equations \ref{eq:prob} to \ref{eq:con} illustrate the calculation of probability, support, and confidence criteria, while Equations \ref{eq:com} and \ref{eq:exp} demonstrate the CPIR calculation process for generating exception rules, where $\neg Y$ denotes the rule is exception rule. 

Additionally, to facilitate the comparison of rule generation efficiency across different ARM algorithms, this study incorporated running time for measuring efficiency.

\begin{equation}\label{eq:px}
    P_X = \frac{freq(X_n^m)}{N}
\end{equation}

\begin{equation}\label{eq:prob}
    P = \prod_{i=1}^{m}p(Y_{n}^{i}|X_{n}^{i})
\end{equation}

\begin{equation}\label{eq:sup}
    Support = \frac{freq(X \Rightarrow Y)}{N}
\end{equation}

\begin{equation}\label{eq:con}
    Confidence = \frac{freq(X \Rightarrow Y)}{freq(X)}
\end{equation}

\begin{equation}\label{eq:com}
    CPIR(X \Rightarrow Y) = \frac{sup(X \bigcup Y) - sup(X) \times sup(Y)}{sup(X) \times (1-sup(Y))}
\end{equation}

\begin{equation}\label{eq:exp}
    CPIR(X \Rightarrow \neg Y) = \frac{sup(X \bigcup \neg Y) - sup(X) \times sup(\neg Y)}{sup(X) \times sup(Y)}
\end{equation}


\subsection{Datasets Descriptions}

This study deployed two open-access databases from the UC Irvin machine learning repository as illustrative case study.
The adopted datasets represent the LMS application in high school and higher education to evaluate the proposed mining module while ensuring the reproducibility of our work.

\subsubsection{Dataset-I High School Student Performance Dataset} 
Dataset-I was proposed by Cortez and Silva \cite{Cor2014}, which archives student performance from two high schools. 
The initial dataset consists of two separate datasets with $33$ attributes, including students' demographics, societal features, and final grades in math and Portuguese, which were all collected through school reports and questionnaires.
During the pre-processing stage, the two datasets were merged into a single dataset, and the duplicates were eliminated.  
The resulting dataset contains $1,044$ instances with $28$ selected attributes.
More detailed information can be found in Table \ref{tab:1}.

\begin{table}[!t]
\caption{Dataset-I Description\label{tab:1}}
\centering
\begin{tabular}{|p{3cm}||p{5cm}|}
\hline
{\bf Attributes} & {\bf Information} \\
\hline
Sex & Female, Male \\
Age & $<18$, $\geq18$ \\
Family Size & $\leq3$, $\geq3$ \\
Parent cohabitation status & Living together, Apart \\
Mother's education & None, Primary, $5-9$th, Secondary, Higher \\
Father's education & No, Primary, $5-9$th, Secondary, Higher \\
Mother's job & Teacher, Health, Civil, Home, Other \\
Father's job & Teacher, Health, Civil, Home, Other \\
Reason & Close, Reputation, Course, Other \\
Student's guardian & Mother, Father, Other \\
Travel time & $<15$min, $15-30$, $30$min-$1$h, $>1$h \\
Weekly study time & $<2$h, $2-5$h, $5-10$h, $>10$h \\
Number of failures & $0-4$ times \\
Extra educational support & True, False \\
Family support & True, False \\
Extra paid classes & True, False \\
Extra curricular activities & True, False \\
Attended nursery school & True, False \\
Take higher education & True, False \\
Internet access at home & True, False \\
With a relationship & True, False \\
Quality of relationships & Very bad to excellent \\
Free time after school & Very low to very high \\
Going out with friends & Very low to very high \\
Workday alcohol consume & Very low to very high \\
Weekend alcohol consume & Very low to very high \\
Current health status & Very bad to very good \\
Number of absences & From $0$ to $93$ \\
\hline
Final Grade (G$3$) & Bad, Average, Good \\
\hline
\end{tabular}
\end{table}

\subsubsection{Dataset-II Higher Education Student Performance Dataset} 
Dataset-II was collected in $2019$ at the Faculty of Engineering and Faculty of Educational Sciences from Near East University \cite{Yil2019}. 
This dataset consists of $32$ attributes that cover personal, family, and educational-related features of the university students. 
After pre-processing, the dataset includes $147$ instances with $28$ attributes.
More detailed information can be found in Table \ref{tab:2}.

\begin{table}[!t]
\caption{Dataset-II Description\label{tab:2}}
\centering
\begin{tabular}{|p{3cm}||p{5cm}|}
\hline
{\bf Attributes} & {\bf Information} \\
\hline
Sex & Female, Male \\
Age & $18-21$, $22-25$, $\geq26$ \\
Graduated high-school & Private, State, Other \\
Scholarship type & None, $25\%$, $50\%$, $75\%$, Full \\
Additional work & True, False \\
Regular activity & True, False \\
Partner & True, False \\
Total salary & $135-200$, $201-270$, $271-340$, $341-410$, $>410$ USD \\
Transportation & Bus, Private car/taxi, Bicycle, Other \\
Accommodation & Rental, Dormitory, With family, Other \\
Mother education & Primary, Secondary, High, University, MSc, Ph.D. \\
Father education & Primary, Secondary, High, University, MSc, Ph.D. \\
Number of siblings & From $1$ to $5$, $>5$ \\ 
Parent status & Married, Divorced, Died \\
Mother occupation & Retired, Housewife, Officer, Private sector employee, Self-employment, Other \\
Father occupation & Retired, Housewife, Officer, Private sector employee, Self-employment, Other \\
Weekly study hours & None, $<5$h, $6-10$h, $11-20$h, $\geq20$h \\
Reading frequency (non-scientific) & None, Sometimes, Often \\
Reading frequency (scientific) & None, Sometimes, Often \\
Attendance to seminars & True, False \\ 
Impact of projects on success & Positive, Negative, Neutral \\
Attendance to classes & Always, Sometimes, Never \\
Prepare exams & Alone, With friends, Not applicable \\
Prepare exams time & Closest date to the exam, Regularly during the semester, Never \\
Taking notes in classes & Never, Sometimes, Always \\
Listening in classes & Never, Sometimes, Always \\
Discussion for success & Never, Sometimes, Always \\
Flip-classroom & Not useful, Useful, Not applicable \\
\hline
Final Grade & Bad, Average, Good \\
\hline
\end{tabular}
\end{table}

\subsection{Parameters Setting}
As those data were recorded in the LMS in different formats, a rigorous pre-processing step is required before inputting them into the ARM module.
In this case, the LMS recorded dataset will need to follow the below mechanism during the experimental setup before applying the ARM module:
\begin{itemize}
    \item All the instances with missing variables should be removed.
    \item All the categorical values should be transformed into pre-defined classes.
    \item Make sure there are no duplicate values for each categories.
    \item Numerical values should be categorized by assigning group intervals.
\end{itemize}

Additionally, the threshold values were pre-defined during the ARM implementation through fine-tuning.
Notably, these thresholds should be adjusted based on the stored dataset information, such as its size and dimensionality.
In this case, we pre-define the support threshold which was $0.7$, and the confidence threshold was set as $0.9$ for extracting common rules.
As for generating the exception rules, the pre-defined support interval was $(0.2, 0.4]$ (i.e., $>$ $0.2$ and $\leq$ $0.4$).
Additionally, the confidence threshold was set as the same for generating common rules.
The CPIR threshold was set to $0.6$ for both datasets.

\section{Results}
This section interprets the extracted association rules from the involved datasets. 
The student performance is further evaluated through a classification task using the proposed algorithm and feature selection techniques for better comparison.

\subsection{ARM Results}
This section presents the extracted rules, along with a performance comparison of the selected ARM algorithms to comprehensively evaluate the proposed method.

Notably, the results generated by the proposed Faster Apriori algorithm were compared to the classic Apriori process concerning confidence and efficiency.
After careful comparisons, the Faster Apriori algorithm's generated rules were comparatively similar to those generated by Apriori.
Therefore, the knowledge extraction of the Faster Apriori algorithm was efficiently improved without information loss.

\subsubsection{Extracted Association Rules}
The proposed algorithm was applied to extract association rules, and the results can be found in Table \ref{tab:3} and Table \ref{tab:4}.
The generated rules were confirmed by the classic Apriori algorithm, where only the mutual rules were maintained and interpreted.

\begin{table*}[!t]
\caption{Dataset-I Extracted Association Rules\label{tab:3}}
\centering
\begin{tabular}{|p{11cm}cc||c|c|c|}
\hline
{\bf Association Rules} & & {\bf Class} & \textbf{Confidence} & \textbf{CPIR} \\
\hline
Male, Family\_Size$\geq3$, Parent\_status=Together & $\Rightarrow$ & \multirow{5}{*}{Final\_grade = Bad} & $1.00$ & $1.00$ \\
Weekend\_alcohol=Low, Workday\_alcohol=Low & $\Rightarrow$ & & $0.98$ & $0.95$ \\
Parent\_status=Together, Family\_Size$\geq3$, Curricular\_activities=T & $\Rightarrow$ & & $0.97$ & $0.74$ \\
Male, Parent\_status=T, Internet=T & $\Rightarrow$ & & $0.97$ & $0.69$ \\
Age$<18$, Weekly\_study\_time=$2-5h$, Higher\_education\_desire=T & $\Rightarrow$ & & $0.96$ & $0.78$ \\
\hline
Mother\_education=Higher, Failures=$0$, Higher\_education\_desire=T & $\Rightarrow$ & \multirow{5}{*}{Final\_grade = Average} & $1.00$ & $1.00$ \\
Age$<18$, Family\_support=T, Failures=$0$, Travel\_time$<15min$, Higher\_education\_desire=T & $\Rightarrow$ & & $1.00$ & $1.00$ \\
Attend\_Nursery=T, Weekend\_alcohol=Low, Workday\_alcohol=Low & $\Rightarrow$ & & $0.99$ & $0.98$ \\
Parent\_status=Together, Weekend\_alcohol=Low, Workday\_alcohol=Low & $\Rightarrow$ & & $0.99$ & $0.96$ \\
Family\_support=T, Guardian=Mother, Failures=$0$, Attend\_Nursery=T & $\Rightarrow$ & & $0.99$ & $0.92$ \\
\hline
Parent\_status=Together, Internet=T, Failures=$0$, Attend\_Nursery=T, Higher\_education\_desire=T, Mother\_education=Higher, Travel\_time$<15min$ & \multirow{2}{*}{$\Rightarrow$} & \multirow{7}{*}{Final\_grade = Good} & \multirow{2}{*}{$1.00$} & \multirow{2}{*}{$1.00$} \\
Age$<18$, Parent\_status=Together, Guardian=Mother, Attend\_Nursery=T, Failures=$0$, Weekend\_alcohol=Low & \multirow{2}{*}{$\Rightarrow$} & & \multirow{2}{*}{$1.00$} & \multirow{2}{*}{$1.00$} \\
Internet=T, Attend\_Nursery=T, Workday\_alcohol=Low & $\Rightarrow$ & & $0.98$ & $0.92$ \\
Failures=$0$, Higher\_education\_desire=T, Weekend\_alcohol=Low, Workday\_alcohol=Low & $\Rightarrow$ & & $0.98$ & $0.90$ \\ 
Age$<18$, Failures=$0$, Weekend\_alcohol=Low, Workday\_alcohol=Low & $\Rightarrow$ & & $0.98$ & $0.90$ \\
\hline
\end{tabular}
\end{table*}

Table \ref{tab:3} presents the extracted common and exception association rules from Dataset-I. 
It should be noted that students' performance is categorized into three classes: performing poorly (``bad'' class with scores ranging from $0$ to $10$), normally (``average'' class with scores ranging from $10$ to $15$), and well (``good'' class with scores ranging from $15$ to $20$). 
It can be observed that male students from families with more than three members and parents living together tend to perform poorly in their final exams, with a confidence and CPIR value of $1.00$. 
Alcohol consumption does not appear to have a significant impact on student exam performance, while participation in extra-curricular activities may be influential, with a confidence value of $0.97$ and a CPIR value of $0.74$. 
Furthermore, limited weekly study time and Internet access at home are associated with poor performance.
On the other hand, students with parents living together, Internet access at home, without past exam failures, have higher education desire, mother has a higher education degree, with a travel time less than $15$ minutes tend to perform well on their final exams.
Notably, the absence of past exam failures appears to be a critical factor in the well-performing student group.

\begin{table*}[!t]
\caption{Dataset-II Extracted Association Rules\label{tab:4}}
\centering
\begin{tabular}{|p{11cm}cc||c|c|c|}
\hline
{\bf Association Rules} & & {\bf Class} & \textbf{Confidence} & \textbf{CPIR} \\
\hline
Salary=$135-200USD$, Class\_attendance=Always & $\Rightarrow$ & \multirow{5}{*}{Final\_grade = Bad} & $1.00$ & $1.00$ \\
Salary=$135-200USD$, Preparation\_date=Close\_to\_exam & $\Rightarrow$ & & $1.00$ & $1.00$ \\
Age=$18-21$, Mother\_education=High\_school	& $\Rightarrow$ & & $1.00$ & $1.00$ \\
Weekly\_study\_time$<5h$, Preparation\_date=Close\_to\_exam & $\Rightarrow$ & & $1.00$ & $1.00$ \\
Weekly\_study\_time$<5h$, Partner=T & $\Rightarrow$ & & $1.00$ & $1.00$ \\
\hline
Parent\_status=Married, Take\_notes=Always, Discussion=Always & $\Rightarrow$ & \multirow{7}{*}{Final\_grade = Average} & $1.00$ & $1.00$ \\
Male, Parent\_status=Married, Class\_attendance=Always, Seminar\_attendance=T & $\Rightarrow$ & & $1.00$ & $1.00$ \\
Parent\_status=Married, Mother\_education=Primary\_school, Mother\_occupation=Housewife, Transportation=Bus & \multirow{2}{*}{$\Rightarrow$} & & \multirow{2}{*}{$1.00$} & \multirow{2}{*}{$1.00$} \\
Male, Parent\_status=Married, Mother\_occupation=Housewife, Class\_attendance=Always, Seminar\_attendance=T & \multirow{2}{*}{$\Rightarrow$} & & \multirow{2}{*}{$1.00$} & \multirow{2}{*}{$1.00$} \\
Male, Parent\_status=Married, Project\_on\_success=Positive, Seminar\_attendance=T & $\Rightarrow$ & & $1.00$ & $1.00$ \\
\hline
Male, Take\_notes=Always, Project\_on\_success=Positive & $\Rightarrow$ & \multirow{7}{*}{Final\_grade = Good} & $1.00$ & $1.00$ \\
Male, Project\_on\_success=Positive, Class\_attendance=Always, Listening\_in\_class=Sometimes & $\Rightarrow$ & & $1.00$ & $1.00$ \\
Parent\_status=Married, Preparation\_for\_exam=Alone, Preparation\_date=Close\_to\_exam, Seminar\_attendance=T & \multirow{2}{*}{$\Rightarrow$} & & \multirow{2}{*}{$1.00$} & \multirow{2}{*}{$1.00$} \\
Parent\_status=Married, Class\_attendance=Always, Seminar\_attendance=T, Preparation\_for\_exam=Alone, Preparation\_date=Close\_to\_exam & \multirow{2}{*}{$\Rightarrow$} & & \multirow{2}{*}{$1.00$} & \multirow{2}{*}{$1.00$} \\
Parent\_status=Married, Mother\_occupation=Housewife, Graduated\_school=State & $\Rightarrow$ & & $1.00$ & $1.00$ \\
\hline
\end{tabular}
\end{table*}

Table \ref{tab:4} showcases the extracted common and exception association rules from Dataset-II. 
Notably, it is intriguing to observe that students who have an additional job earning a salary ranging from $135$ to $200$ dollars tend to perform poorly in their final exams, despite their regular class attendance. 
This rule exhibits both a high confidence and CPIR value of $1.00$. 
Additionally, students with a weekly study time of less than $5$ hours are strongly associated with poor performance.
Conversely, students who have parents with a married status, prepare for exams alone, and attend extra seminars tend to perform well in their final exams. Furthermore, being male, having a high class attendance rate, demonstrating a positive impact on previous projects, preparing for exams close to the date, and graduating from state schools are all positively related factors to the well-performed exam.

\subsubsection{ARM Efficiency Comparison}
This section compares the running time among the ARM techniques, including the classic Apriori, FP-Tree, and the proposed Faster Apriori algorithm.

\begin{figure}[!t]
\centering
\includegraphics[width=1\columnwidth]{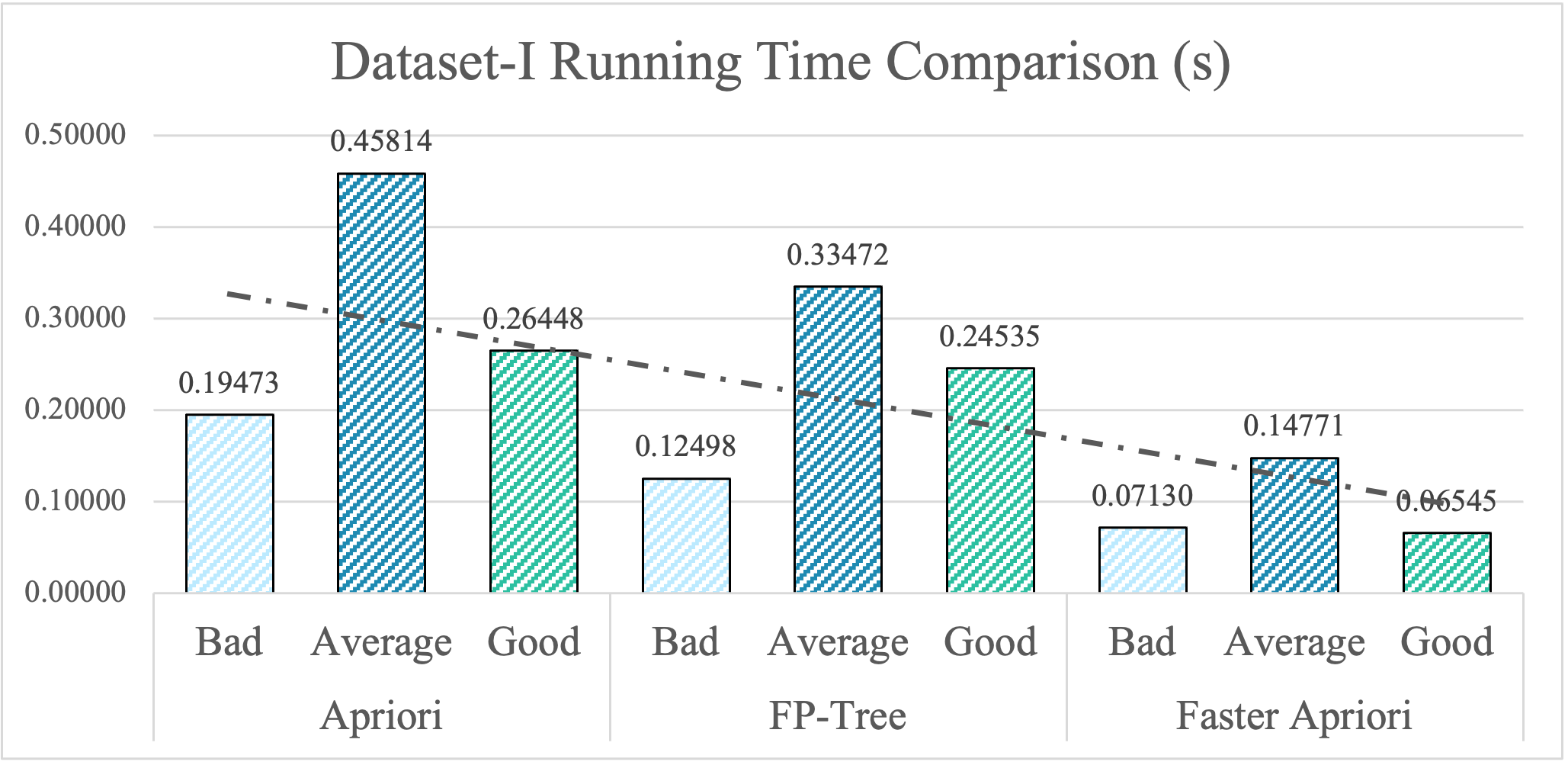}
\caption{Dataset-I Running Time Comparison.}
\label{fig3}
\end{figure}

\begin{figure}[!t]
\centering
\includegraphics[width=1\columnwidth]{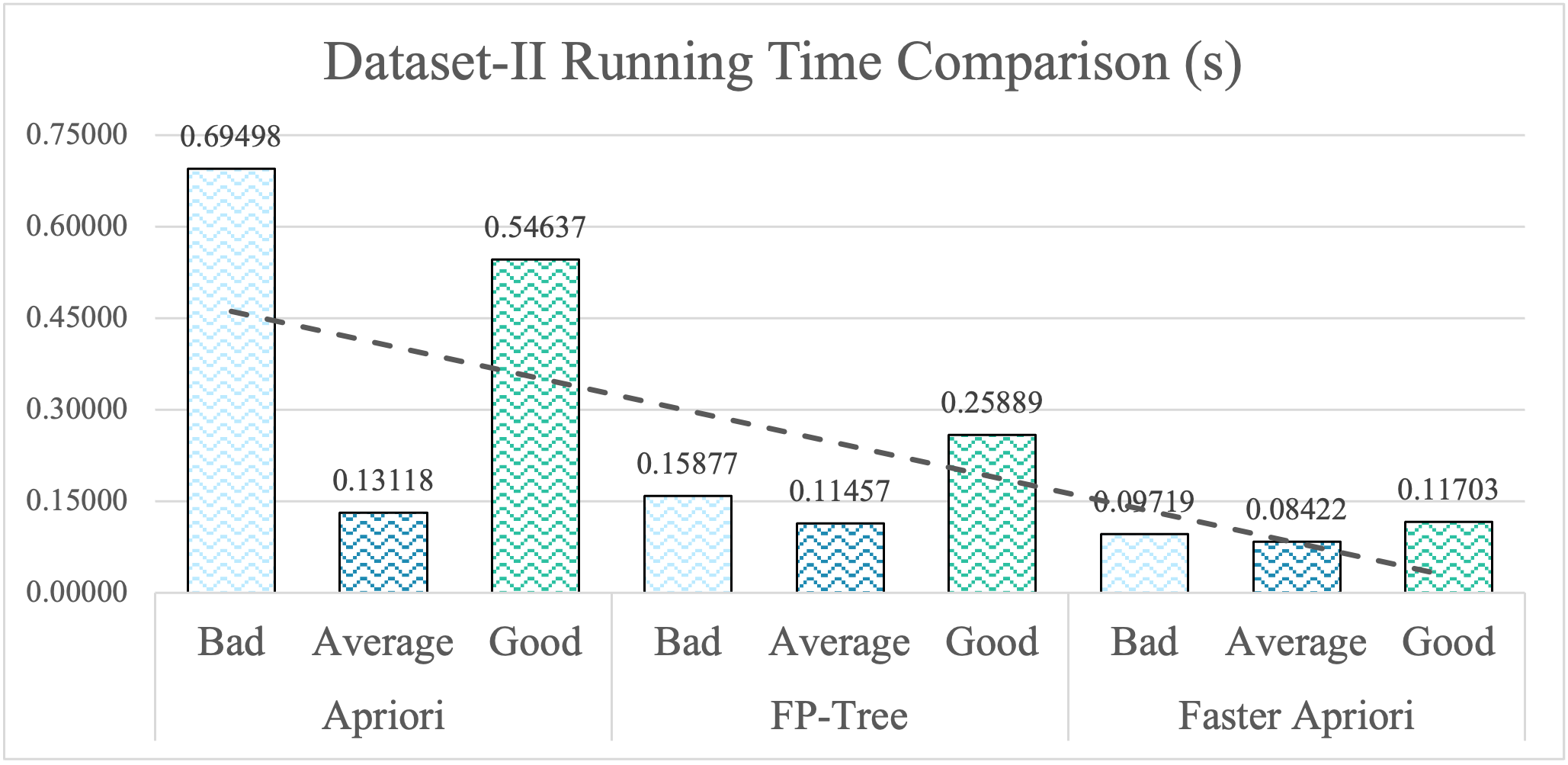}
\caption{Dataset-II Running Time Comparison.}
\label{fig4}
\end{figure}

Figures \ref{fig3} and \ref{fig4} depict the running time comparison for the two independent datasets.
It should be noted that the running time combines the time for frequent itemsets identification and association rules generation.
Each of the algorithmic running time was measured based on different classes.
More specifically, the Apriori algorithm produces an average running time of $0.306$ seconds for Dataset-I, the FP-Tree reduced the running time to $0.235s$, whereas the proposed faster Apriori took $0.095s$ for common and exception rules generation.
Similarly, the Apriori, FP-Tree, and Faster Apriori algorithms generated rules with $0.458s$, $0.177s$, and $0.099s$ for Dataset-II, showing significantly improved efficiency.

Based on the analysis, the time and space complexity of the classic Apriori algorithm is Big O = $\mathcal{O}(2^d)$.
In this case, $d$ is the number of unique factors in the given database $D$.
The algorithm needs to scan through the database to identify frequent item sets during each iteration, which increases the running time following an exponential distribution.
As the run-time grows exponentially with the size of the input, this study delivers the Faster Apriori algorithm to update the original database during each running iteration.
Under such a condition, the proposed Faster Apriori algorithm has a time complexity of Big O = $\mathcal{O}((m+k) \times n)$.
Specifically, $m$ is the number of frequent itemsets, $k$ is the number of iterations performed until no more frequent itemsets are found, and 
$n$ is the number of transactions in the dataset $\mathcal{D}$.
The space complexity of Faster Apriori is Big O = $\mathcal{O}(n)$, which depends on the largest data structure in the database.
$\mathcal{O}((m+k) \times n)$ represents a polynomial time complexity, which is more desirable than exponential time complexity $\mathcal{O}(2^d)$ when dealing with larger datasets or input volumes. 
Accordingly, the database becomes smaller in scale in each iteration, leading to increased efficiency and reduced computational costs of the proposed algorithm compared to the classic Apriori algorithm.

\subsection{Performance Prediction}
With the objective to evaluate the proposed Faster Apriori algorithm, this section includes a classification task to predict student's academic performance.
In doing so, a series of classic classification algorithms were involved: Logistic Regression (\emph{LR}), Decision Tree (\emph{DT}), Support Vector Machine (\emph{SVM}), Random Forest (\emph{RF}), Naive Bayes (\emph{NB}), and Multi-layer Perceptron (\emph{MLP}).

\begin{figure}[!t]
\centering
\includegraphics[width=1\columnwidth]{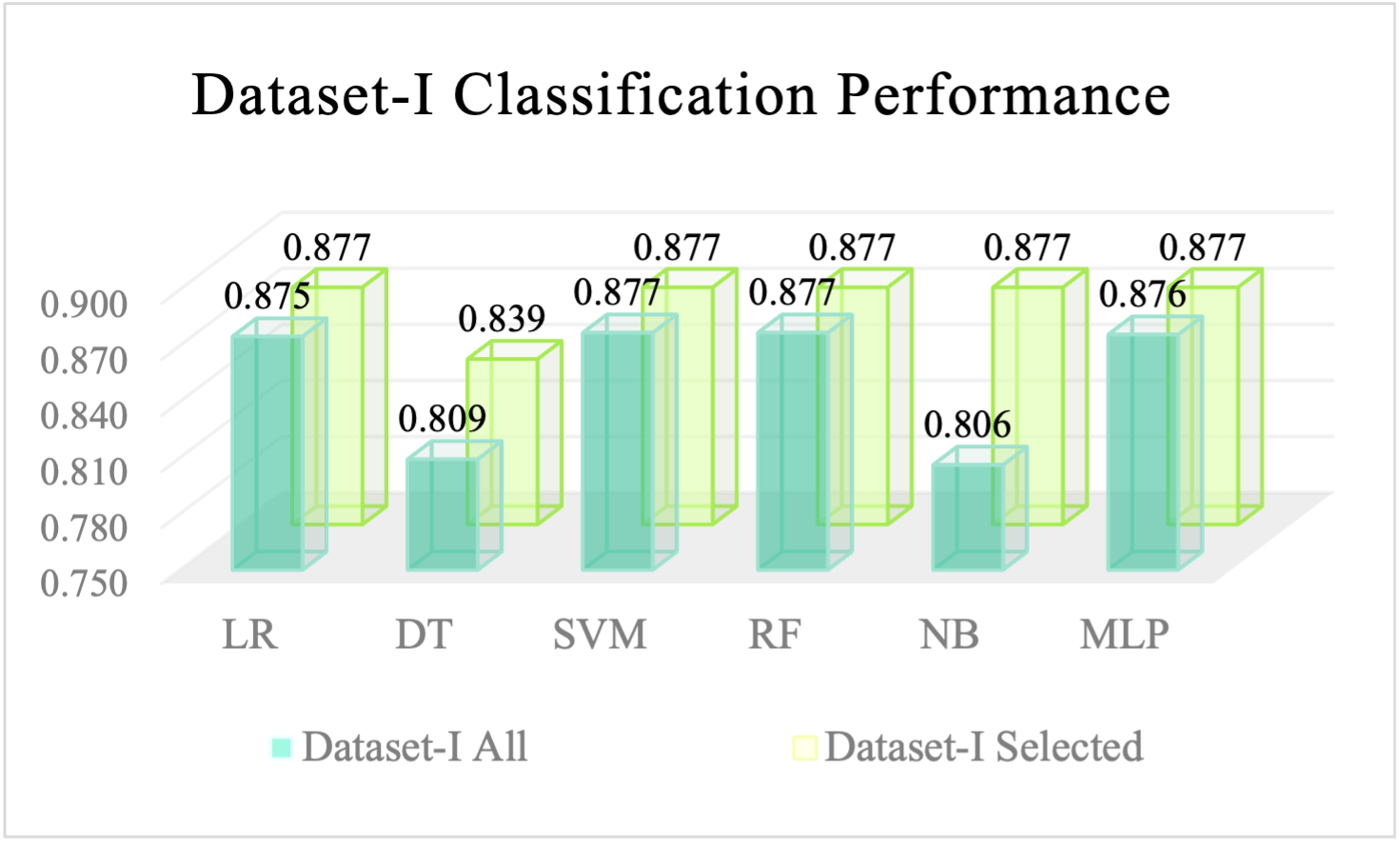}
\caption{Dataset-I Classification Comparison.}
\label{fig5}
\end{figure}

\begin{figure}[!t]
\centering
\includegraphics[width=1\columnwidth]{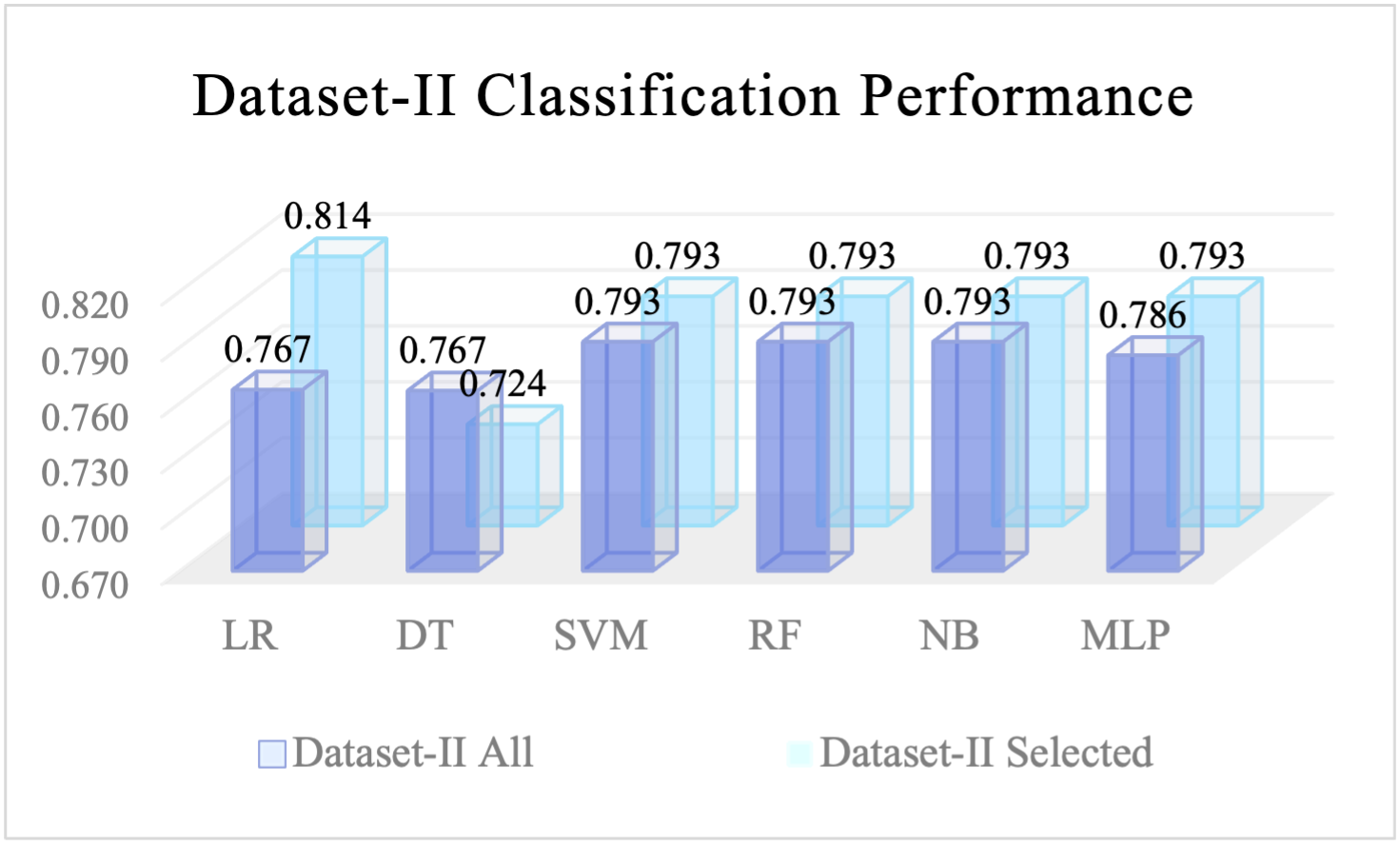}
\caption{Dataset-II Classification Comparison.}
\label{fig6}
\end{figure}

Figures \ref{fig5} and \ref{fig6} illustrate the predictive results of student performance using the two datasets before and after the ARM-based feature selection procedure. 
The proposed algorithm effectively generates strongly associated rules, indicating the impact of the antecedents on the consequence. 
These significant attributes influencing student performance were considered critical attributes, thus, were selected for further evaluation.
For Dataset-I, the proposed Faster Apriori algorithm selected a total of $16$ attributes, including sex, age, family size, parental cohabitation status, mother's education, workday and weekend alcohol consumption, curricular activities, Internet access, weekly study time, higher education desire, failures, family support, travel time, nursery attendance, and guardian.
Similarly, for Dataset-II, attributes such as sex, age, salary, graduate school type, class attendance, mother's education, mother's occupation, weekly study time, partner, parental status, notes-taking, seminar attendance, transportation, project on success, preparation date and type, listening, and discussion were selected.
These selected attributes provide valuable insights into the factors influencing student performance and are deemed significant for predictive analysis.

The performance of the Faster Apriori algorithm-based feature selection technique was evaluated using the $10$-fold cross-validation (\emph{CV}) technique, and the results were compared to the performance of the original datasets without feature selection. 
The evaluation revealed notable improvements in performance after feature selection for both datasets.
For Dataset-I, significant accuracy improvements were observed in LR, DT, NB, and MLP algorithms, while SVM and RF maintained their high accuracy rates of $0.877$. 
For Dataset-II, the original dataset achieved the highest accuracy of $0.793$ using SVM, RF, and NB. 
However, after feature selection with the LR algorithm, an enhanced accuracy of $0.814$ was achieved.
These results indicate that the proposed Faster Apriori algorithm efficiently generates common and exception rules and can also be applied for feature selection.
With fewer selected attributes, the algorithm achieves competitive and even superior performance, demonstrating its effectiveness in knowledge discovery and improving predictive accuracy.

\section{Discussion}
In this study, an efficient LMS framework incorporating an advanced educational data mining technique is proposed to unravel latent knowledge by exploring the associations between students' learning characteristics and their behaviors. 
The results revealed a series of significant attributes that have a significant impact on student performance. 
Through the analysis of both datasets, mutual attributes such as sex, age, parental cohabitation status, mother's education, weekly study time, and participation in extra-curricular or seminar activities were identified. 
These factors were found to be highly influential in determining students' academic success, which aligns with findings from previous studies \cite{Put2016, Abu2019, Hass2021}. 
This finding suggests that the leading factor for student performance success should be a stable and supporting family.

In addition to the previously mentioned factors, other variables such as family size and support, Internet access at home, guardian, preparation for exams, and discussion have also been determined to be important in influencing student performance. 
Among these factors, active engagement in class discussions and interaction with peers and teachers have been found to strongly correlate with academic success \cite{Alh2020, Get2020}. 
These factors highlight the importance of self-learning, self-control, and self-motivation.
Such an ideal way of learning can be achieved through personalized goal-setting, self-instructions, self-monitoring, and self-reinforcement.
Accordingly, educators can provide more guidance on establishing protocols to drive students' enthusiasm in learning from their own and producing feedback iteratively to progress. 

With the proposed educational data modeling technique, the generated results provide valuable insights for educators in planning future learning pedagogies and improving educational curriculum design \cite{Roja2019}. 
For instance, educators can incorporate course-related activities before class to stimulate critical thinking and encourage discussion. 
Providing platforms for students to interact during and after class, as well as conducting regular consultation activities, can deepen students' understanding of course difficulties and enable them to master both foundational knowledge and application skills effectively.
More importantly, these findings can guide academic planners in improving education quality and reducing surveillance failure rates. 
Moreover, the results can also aid educators involved in the admission process by helping them plan effective admission promotion strategies.

\section{Conclusion}
This work addresses two key challenges in educational data mining for LMS development: the neglect of exceptionality in association rule mining tasks, and the need for efficient handling of large volumes of complex digital records. 
To tackle these challenges, an effective LMS incorporating the Faster Apriori algorithm is proposed, capable of simultaneously extracting both common and exception rules, thereby enabling valuable knowledge discovery. 
The algorithm was applied on two educational datasets to uncover patterns related to student performance. 
A series of critical attributes were identified, demonstrating their high influence on students' academic success. 
Further evaluations were conducted to confirm the significance of these selected attributes.

The Faster Apriori algorithm was delivered as a resource for the community to benchmark existing ARM techniques.
The algorithm used Apriori as a baseline model, and it adopted the probability-based module to shrink the original database before candidate itemsets generation in iteration.
Additionally, the algorithm repelled under-qualified items during each iteration, contributing to increased efficiency.

However, this study also has certain limitations, particularly concerning the quality of data attributes. 
This study requires pre-processing steps to change all input attributes into categorical or discretized variables. 
Integrating both categorical and numerical variables is crucial for generating more precise rules. 
Hence, in future research, we aim to develop a more robust data mining algorithm capable of handling heterogeneous data structures by combining categorical and numerical attributes. 
This approach will enhance the algorithm's application and effectiveness in real-world practical scenarios.

\bibliographystyle{IEEEtran}

\begin{IEEEbiographynophoto}{Xinyu Zhang}
(Member, IEEE) received her Bachelor's degree from the Faculty of Information Technology, Monash University, Melbourne, Victoria, Australia, in $2018$, an Honours degree in Information Technology in $2019$, and a Ph.D. degree in Data Science and Artificial Intelligence from Monash University in $2023$.
Dr. Zhang is currently affiliated with the School of Electronics and Information, Northwestern Polytechnical University, Xi’an, China.
Her research interests include machine learning-driven applications in digital health, education, and brain intelligence.
\end{IEEEbiographynophoto}
\begin{IEEEbiographynophoto}{Vincent CS Lee} 
(Senior Member, IEEE) received BEng (EEE) and MSc(CSE) from National University of Singapore and Australia Federal Government scholarship to pursue PhD from $1988$ through to $1991$ at The University of New Castle, NSW, in Australia. 
From $1973$ to $1974$, Dr. Lee was awarded a joint research scholarship by Ministry of Defence (Singapore) and Ministry of Defence (UK) for postgraduate study at Royal Air Force College, UK in aircraft electrical and instrument systems. 
Dr. Lee was a visiting academic to Tsinghua University in Beijing at the School of Economics and Management from Nov $2006$ to March $2007$, and was also a Visiting Professor to Information Communication Institute of Nanyang Technological University Singapore from July $1994$ through June $1995$.
Since June $2002$, Dr. Lee is an Associate Professor at Machine learning and Deep Learning Discipline of the Department of Data Science and Artificial Intelligence, Faculty of IT, Monash University, Australia. 
Dr. Lee's research interests focus on educational data mining, signal and information processing, AI, vision and language, computer vision and object tracking, and financial technology.
\end{IEEEbiographynophoto}
\begin{IEEEbiographynophoto}{Duo Xu} 
received her Bachelor's degree in English literature from the Chengdu College of Arts and Sciences , China, in $2020$. 
Duo Xu is currently affiliated with the Guangyuan Foreign Language School, Sichuan, China. 
Her research interests include English language education and psychological education in English.
\end{IEEEbiographynophoto}
\begin{IEEEbiographynophoto}{Jun Chen} 
(Member, IEEE) received the B.S., M.S., and Ph.D. degrees in system engineering from Northwestern Polytechnical University, China, in $2001$, $2005$, and $2009$, respectively. 
Since $2012$, he has been an Associate Professor with the School of Electronics and Information, Northwestern Polytechnical University. 
His current research interests include modeling and application based on fuzzy cognitive map, and intelligent decision-making for complex systems.
\end{IEEEbiographynophoto}

\begin{IEEEbiographynophoto}{Mohammad S. Obaidat} 
(Life Fellow of IEEE, Fellow of SCS and Fellow of AAIA) is an internationally known academic/researcher/scientist/scholar. He received his Ph.D. degree in Computer Engineering with a minor in Computer Science from Ohio State University, USA. 
He has received extensive research funding and published To Date ($2023$) over $1,200$ refereed technical articles-About half of them are journal articles, over $100$ books, and over $70$ Book Chapters. 
He is Editor-in-Chief of $3$ scholarly journals and an editor of many other international journals.  
He is founder/co-founder of $5$ IEEE International Conferences. 
Among his previous positions are Advisor to the President of Philadelphia University,  President \& Chair of Board of Directors of SCS, Dean of the College of Engineering at Prince Sultan University, Founding Dean of College of Computing and Informatics at University of Sharjah, Chair and tenured Professor at Department of Computer \& Information Science and Director of Graduate Program in Data Analytics at Fordham university, Chair and tenured Professor of the Department of Computer Science and Director of Graduate Program at Monmouth University, Chair and Professor of Computer Science Department at University of Texas-Permian Basin, Distinguished Professor at IIT-Dhanbad, Tenured Full Professor at King Abdullah II School of Information Technology (KASIT), University of Jordan, The PR of China Ministry of Education Distinguished Overseas Professor at the University of Science and Technology Beijing, China and an Honorary Distinguished Professor at the Amity University- A Global University.
He is now a Distinguished Professor at KASIT, University of Jordan.
He has chaired Over $185$ international conferences and has given Over $185$ keynote speeches worldwide. 
He received many best paper awards for his papers in International journals and conferences.
He  received many other worldwide awards for his technical contributions including: $2018$ IEEE ComSoc-Technical Committee on Communications Software $2018$ Technical Achievement Award for contribution to Cybersecurity, Wireless Networks Computer Networks and Modeling and Simulation, SCS prestigious McLeod Founder's Award, Presidential Service Award, SCS Hall of Fame –Lifetime Achievement Award, SCS Outstanding Service Award, Nokia Distinguished Fellowship Award and Fulbright Distinguished Scholar Award, among many others.
Springer published in $2022$ a book honoring his contributions entitled: Advances in Computing, Informatics, Networking and Cybersecurity - A Book Honoring Professor Mohammad S. Obaidat’s Significant Scientific Contributions. 
\end{IEEEbiographynophoto}

\vfill

\end{document}